\documentclass[12pt]{article}             
\begin{document}
\begin{center}
{\bf New inequality for density matrices of single qudit states}\\
V. N. Chernega, O.~V.~Man'ko, V. I. Man'ko\\
 P.N.~Lebedev Physical Institute, Russian Academy of Sciences\\
      Leninskii Prospect, 53, Moscow 119991, Russia \\
      Email: omanko@sci.lebedev.ru
\end{center}
\begin{abstract}
Using the monotonity of relative entropy of composite quantum systems we obtain new entropic inequalities for arbitrary density matrices of single qudit states. Example of qutrit state inequalities and the "qubit portrait" bound for the distance between the qutrit states are considered in explicit form.
\end{abstract}
\vskip1cm

 \vspace{0.4cm}

\section{Introduction}
There exist different entropic inequalities for composite quantum systems \cite{NielsonCH,LiebRuskai,PetzNielson,CorlenLieb2007,20pr,2014,fromRusk,Ruskai}. Recently it was formulated \cite{OlgaVova,OlgaVova1,RitaPS2014,maps,OlgaVovaVIminkovski} that the inequalities known for composite systems uncluding subadditivity condition and strong subadditivity condition can be generalized for noncomposite systems like single qudit. The aim of this work is to use the developed approach \cite{OlgaVova,OlgaVova1,RitaPS2014,maps,OlgaVovaVIminkovski} and to get the new entropic inequalities for single qudit states on the base of known properties of relative entropy of composite systems like the monotonicity of the relative entropy. We obtain a generic new matrix inequality for Hermitian matrices. In case where the matrices coincide with the two density matrices of bipartite quantum system states the new inequality coincides with the monotonicity property of the relative entropy. But it is valid for arbitrary pair of density matrices also. The bound for distance between two arbitrary density matrices is obtained and expressed in terms of matrix elements of these matrices. The paper is organized as follows. In next Sec. 2 the property of relative entropy are considered. The example of qutrit and 4$\times4$ - density matrices are presented in Sec. 3. Conclusion and perspectives are given in Sec. 4.

\section{Relative entropy}
The relative entropy between density matrices $\rho$ and $\sigma$ is defined as 
\begin{equation}\label{NEADM1}
S(\rho||\sigma)=\mbox{Tr}(\rho\ln\rho-\rho\ln\sigma),
\end{equation}
where $\rho$ and $\sigma$ are density matrices of quantum states. Also if the density matrices are the density matrices of bipartite system states, i.e. $\rho\mapsto\rho(1,2)$ and $\sigma\mapsto\sigma(1,2)$, one has inequality
\begin{equation}\label{NEADM2}
\mbox{Tr}\left(\rho(1,2)\ln\rho(1,2)-\rho(1,2)\ln\sigma(1,2)\right)\geq\mbox{Tr}\left(\rho(1)\ln\rho(1)-\rho(1)\ln\sigma(1)\right),
\end{equation}
where
\[\rho(1)=\mbox{Tr}_2\,\rho(1,2),\quad \sigma(1)=\mbox{Tr}_2\,\sigma(1,2).\]
This inequality reflects the monotonicity property of relative entropy.

Another analogous inequality holds
\begin{equation}\label{NEADM2}
\mbox{Tr}\left(\rho(1,2)\ln\rho(1,2)-\rho(1,2)\ln\sigma(1,2)\right)\geq
\mbox{Tr}\left(\rho(2)\ln\rho(2)-\rho(2)\ln\sigma(2)\right),
\end{equation}
where
\[\rho(2)=\mbox{Tr}_1\,\rho(1,2),\quad \sigma(2)=\mbox{Tr}_1\,\sigma(1,2).\]
Now we extend the inequalities for arbitrary matrices. For example, given a density $N\times N$ matrix $\rho$ in block form
\[\rho=\left(\begin{array}{cc}
A&B\\C&D\end{array}\right)\]
and two $N\times N$-matrices
\[\rho_A=\left(\begin{array}{cc}
A&0\\0&0\end{array}\right),\quad \rho_D=\left(\begin{array}{cc}
D&0\\0&0\end{array}\right).\]
Let us construct positive map $\rho\mapsto\rho_1$ where
\[\rho_1=\rho_A+\rho_D\equiv\left(\begin{array}{cc}
A+D&0\\0&0\end{array}\right).\]
The block $D$ is $m\times m$-matrix, i.e. block $A$ is $(N-m)\times(N-m)$-matrix and $m\leq N-m$.

The following matrix inequality holds for the density matrix
$\left(\begin{array}{cc}
A&B\\C&D\end{array}\right)$ given in block form
\[\mbox{Tr}\left[\left(\begin{array}{cc}
A+D&0\\0&0\end{array}\right)\ln\left(\begin{array}{cc}
A+D&0\\0&0\end{array}\right)-\left(\begin{array}{cc}
A+D&0\\0&0\end{array}\right)\ln\left(\begin{array}{cc}
A&B\\C&D\end{array}\right)\right]\geq0.\]
Another inequality for this matrix reads
\[\mbox{Tr}\left[\left(\begin{array}{ccc}
\mbox{Tr}\,A&\mbox{Tr}\,B&0\\\mbox{Tr}\,C&\mbox{Tr}\,D&0\\0&0&0\end{array}\right)\ln\left(\begin{array}{ccc}
\mbox{Tr}\,A&\mbox{Tr}\,B&0\\\mbox{Tr}\,C&\mbox{Tr}\,D&0\\0&0&0\end{array}\right)-\left(\begin{array}{ccc}
\mbox{Tr}\,A&\mbox{Tr}\,B&0\\\mbox{Tr}\,C&\mbox{Tr}\,D&0\\0&0&0\end{array}\right)\ln\left(\begin{array}{cc}
A&B\\C&D\end{array}\right)\right]\geq0.\]
This inequality means the nonnegativity condition for relative entropy of the initial density matrix $\left(\begin{array}{cc}
A&B\\C&D\end{array}\right)$ and its "portraits" obtained by applying two positive maps
\[\left(\begin{array}{cc}
A&B\\C&D\end{array}\right)\mapsto\left(\begin{array}{cc}
A+D&0\\0&0\end{array}\right), \quad\left(\begin{array}{cc}
A&B\\C&D\end{array}\right)\mapsto\left(\begin{array}{ccc}
\mbox{Tr}\,A&\mbox{Tr}\,B&0\\\mbox{Tr}\,C&\mbox{Tr}\,D&0\\0&0&0\end{array}\right).\]
The new matrix inequality connected with monotonicity property of relative entropy of two density matrices
\[\rho=\left(\begin{array}{cc}
A&B\\C&D\end{array}\right),\quad \sigma=\left(\begin{array}{cc}
a&b\\c&d\end{array}\right)\]
is given by the relation
\[\mbox{Tr}\left[\left(\begin{array}{cc}
A&B\\C&D\end{array}\right)\ln\left(\begin{array}{cc}
A&B\\C&D\end{array}\right)-\left(\begin{array}{cc}
A&B\\C&D\end{array}\right)\ln\left(\begin{array}{cc}
a&b\\c&d\end{array}\right)\right]\]
\[\geq\mbox{Tr}\left[\left(\begin{array}{cc}
A+D&0\\0&0\end{array}\right)\ln\left(\begin{array}{cc}
A+D&0\\0&0\end{array}\right)-\left(\begin{array}{cc}
A+D&0\\0&0\end{array}\right)\ln\left(\begin{array}{cc}
a+d&0\\0&0\end{array}\right)\right].\]
This inequality holds for two arbitrary nonnegative matrices with $\mbox{Tr}\rho=\mbox{Tr}\sigma=1$.

\section{Qutrit and qudit examples}
The example of qutrit density matrices $\rho$ and $\sigma$ provides the new inequality
\[
\mbox{Tr}\left[\left(
       \begin{array}{ccc}
         \rho_{11} & \rho_{12} & \rho_{13} \\
         \rho_{21} & \rho_{22} & \rho_{23} \\
         \rho_{31} & \rho_{32} & \rho_{33} \\
       \end{array}
     \right)\left[\ln \left(
       \begin{array}{ccc}
         \rho_{11} & \rho_{12} & \rho_{13} \\
         \rho_{21} & \rho_{22} & \rho_{23} \\
         \rho_{31} & \rho_{32} & \rho_{33} \\
       \end{array}
     \right)- \ln \left(
       \begin{array}{ccc}
         \sigma_{11} &\sigma_{12} &\sigma_{13} \\
         \sigma_{21} &\sigma_{22} &\sigma_{23} \\
       \sigma_{31} &\sigma_{32} &\sigma_{33} \\
       \end{array}\right) \right]\right]\geq\]
       \[\mbox{Tr}\left[\left(
         \begin{array}{cc}
           \rho_{11}+\rho_{33} & \rho_{12} \\
           \rho_{21} & \rho_{22}\end{array}\right)\left[\ln\left(
         \begin{array}{cc}
           \rho_{11}+\rho_{33} & \rho_{12} \\
           \rho_{21} & \rho_{22}\end{array}\right)-\ln\left(
         \begin{array}{cc}
           \sigma_{11}+\sigma_{33} & \sigma_{12} \\
           \sigma_{21} & \sigma_{22}\end{array}\right)\right]\right].\]
The relative entropy of two states is known to be nonnegative characteristic of a distance between the states. The new above inequality gives the bound for the distance which equals to relative entropy of two "qubits". The bound equals to the maximal "qubits" distance obtained by all the permutations of indices 1,2,3, in the qutrit states. Thus the problem of distance bound of qudits expressed in terms of the relative entropy is reduced to the problem of distance of "qubit portrait" \cite{Vovf,Lupo} of the qudit states. This property is analogous to the property of entanglement of composite qudit system states which can be characterized by the entanglement of the qubit portrait of these states. One can write the chain of the inequalities for arbitrary density matrices $\rho$ and $\sigma$. We demonstrate such chain for density $4\times4$-matrices $\rho$ and $\sigma$ which can be associated either with two-qubit state or with the states of qudit with $j=3/2$. One has the chain of inequalities
\[
\mbox{Tr}
\left\{
\left(\begin{array}{cccc}
\rho_{11}&\rho_{12}&\rho_{13}&\rho_{14}\\
\rho_{21}&\rho_{22}&\rho_{23}&\rho_{24}\\
\rho_{31}&\rho_{32}&\rho_{33}&\rho_{34}\\
\rho_{41}&\rho_{42}&\rho_{43}&\rho_{44}
\end{array}\right)
\left[
\ln\left(\begin{array}{cccc}
\rho_{11}&\rho_{12}&\rho_{13}&\rho_{14}\\
\rho_{21}&\rho_{22}&\rho_{23}&\rho_{24}\\
\rho_{31}&\rho_{32}&\rho_{33}&\rho_{34}\\
\rho_{41}&\rho_{42}&\rho_{43}&\rho_{44}
\end{array}\right)
-\ln\left(\begin{array}{cccc}
\sigma_{11}&\sigma_{12}&\sigma_{13}&\sigma_{14}\\
\sigma_{21}&\sigma_{22}&\sigma_{23}&\sigma_{24}\\
\sigma_{31}&\sigma_{32}&\sigma_{33}&\sigma_{34}\\
\sigma_{41}&\sigma_{42}&\sigma_{43}&\sigma_{44}
\end{array}\right)
\right]
\right\}
\geq
\]
\[\mbox{Tr}
\left\{
\left(\begin{array}{ccc}
\rho_{11}+\rho_{44}&\rho_{12}&\rho_{13}\\
\rho_{21}&\rho_{22}&\rho_{23}\\
\rho_{31}&\rho_{32}&\rho_{33}
\end{array}\right)
\left[
\ln\left(\begin{array}{ccc}
\rho_{11}+\rho_{44}&\rho_{12}&\rho_{13}\\
\rho_{21}&\rho_{22}&\rho_{23}\\
\rho_{31}&\rho_{32}&\rho_{33}
\end{array}\right)-
\ln\left(\begin{array}{ccc}
\sigma_{11}+\sigma_{44}&\sigma_{12}&\sigma_{13}\\
\sigma_{21}&\sigma_{22}&\sigma_{23}\\
\sigma_{31}&\sigma_{32}&\sigma_{33}
\end{array}\right)
\right]
\right\} \geq
\]
\[
\mbox{Tr}
\left\{
\left(\begin{array}{cc}
\rho_{11}+\rho_{33}+\rho_{44} & \rho_{12} \\
\rho_{21} & \rho_{22}\end{array}\right)
\left[
\ln \left(\begin{array}{cc}
\rho_{11}+\rho_{33}+\rho_{44} & \rho_{12} \\
\rho_{21} & \rho_{22}\end{array}\right)
-\ln\left(\begin{array}{cc}
\sigma_{11}+\sigma_{33}+\sigma_{44} & \sigma_{12} \\
\sigma_{21} & \sigma_{22}\end{array}\right)
\right]
\right\}
\geq 0.
\]
Thus the distance of two-qubits or $j=3/2$ qudit states has the bound determined by relative entropy of "qutrit"  states which also has bound determined by relative entropy of "qubit" states. The obtained inequalities are valid for arbitrary density matrices including the noncomposite system states of a single qudit.

For given finite density matrix $\rho$ and arbitrary Hermitian $N\times N$-matrix $B$ one can get inequality
\[ 0\leq\exp\left[-\mbox{Tr}(\rho\ln\rho)\right]\leq\left[(\mbox{Tr}e^{-B})(\mbox{Tr}e^B)\right]^{1/2}.\]
For $B=\rho\ln\rho$ one has:
\[\exp\left[-\mbox{Tr}(\rho\ln\rho)\right]\leq\{[\mbox{Tr}\exp(-\rho\ln\rho)][\mbox{Tr}\exp(\rho\ln\rho)]\}^{1/2}.\]
If $b_k$ $(k=1,2,\ldots,N)$ are eigenvalues of the matrix $B$ the inequality means that there exists the bound for the sum
\[\sum_{k=1}^N\sum_{j=1}^N\exp(b_k-b_j)\geq N^2.\]
For $N=2$ the inequality provides obvious relation
\[ 2[1\cosh(b_1-b_2)]\geq 4\]
which is saturated for $b_1=b_2$. One can also check that for any set of reals $b_k$, $k=1,2,\ldots,N$ and nonnegative numbers $w_k$ such that $\sum_k w_k=1$ the following inequality holds
\[ \ln\sum_{k=1}^N[\exp(-b_k)+\sum_{k=1}^N(w_k\ln w_k+w_kb_k)\geq0.\]
Thus for $N$ real $b_k$ and probability vector $\vec w=\{w_k\}$ one has the generic inequality. If the qudit state is associated with spin tomogram \cite{DodPLA,OlgaJETP,OlgaBregence} or probability vector $\vec w(u)$ on the unitary group $u$ the obtained inequality for $b_k=-w_k(u)$ provides some uncertainty relation for measurable probabilities $w_k(u)$.

\section*{Conclusion}
To conclude we formulate the main result of our work. Using the "portrait" approach \cite{Vovf,Lupo} to the density matrices of single qudit states the analog of monotonicity property of relative entropy valid for bipartite quantum system was constructed for systems without subsystems. New matrix inequalities for qutrit density matrices were obtained. The chain of matrix inequalities for single qudit states generalizing the chain of the  analogous inequalities for multipartite qudit states was demonstrated on example of qudit $j=3/2$ systems and two-qubit system. The generic inequality for $N$ real numbers and a probability $N$-vector was obtained. Also an "uncertainty relation" for spin tomogram was obtained. The possible applications of the obtained inequalities to analyze quantum correlations in the states of noncomposite systems will be considered in future publication.

\end{document}